\documentclass[12pt,thmsa]{article}
\usepackage{amssymb}

\usepackage{sw20lart}

\input tcilatex
\begin{document}

\title{RETARDING SUB- AND ACCELERATING SUPER-DIFFUSION GOVERNED BY DISTRIBUTED
ORDER FRACTIONAL DIFFUSION EQUATIONS }
\author{A. V. Chechkin$^{1,2}$, R. Gorenflo$^{1},$ and I.M. Sokolov$^{3,4}$ \\
%EndAName
$^1$\textit{Erstes Mathematisches Institut, FB Mathematik and Informatik }\\
\textit{\ Freie Universitaet Berlin, }\\
\textit{\ Arnimalee 3 D-14195 Berlin, GERMANY }\\
$^2$\textit{Institute for Theoretical Physics }\\
\textit{\ National Science Center ''Kharkov Institute of Physics and
Technology '',}\\
\textit{\ Akademicheskaya st.1, 61108 Kharkov, UKRAINE }\\
$^{3}$\textit{Institut fur Physik Humboldt-Universitaet zu Berlin, }\\
\textit{\ Invalidenstrasse 110 D-10115 Berlin, GERMANY }\\
$^{4}$\textit{Theoretische Polymerphysik, Universitaet Freiburg,}\\
\textit{\ Hermann-Herder-Strasse 3, D-79104 Freiburg im Breisgau, GERMANY}}
\maketitle

\begin{abstract}
We propose diffusion-like equations with time and space fractional
derivatives of the distributed order for the kinetic description of
anomalous diffusion and relaxation phenomena, whose diffusion exponent
varies with time and which, correspondingly, can not be viewed as
self-affine random processes possessing a unique Hurst exponent. We prove
the positivity of the solutions of the proposed equations and establish the
relation to the Continuous Time Random Walk theory. We show that the \textit{%
distributed order time fractional diffusion equation} describes the
sub-diffusion random process which is subordinated to the Wiener process and
whose diffusion exponent diminishes in time (\textit{retarding sub-diffusion}%
) leading to \textit{superslow diffusion}, for which the square displacement
grows logarithmically in time. We also demonstrate that the \textit{%
distributed order space fractional diffusion equation} describes
super-diffusion phenomena when the diffusion exponent grows in time (\textit{%
accelerating super-diffusion}).

PACS numbers:

02.50.Ey Stochastic processes;

05.40.-a Fluctuation phenomena, random processes, noise and Brownian motion.
\end{abstract}

\newpage

Recently, kinetic equations with fractional space and time derivatives have
attracted attention as a possible tool for the description of anomalous
diffusion and relaxation phenomena, see, e.g., [MK00], [MLP01], [SZ97],
[Mai96] and references on earlier studies therein. It was also recognized
[HA95], [Com96], [MRGS00], [BMK00] that the fractional kinetic equations may
be viewed as ''hydrodynamic'' (that is, long-time and long-space) limits of
the CTRW (Continuous Time Random Walk) theory [MW65] which was succesfully
applied to the description of anomalous diffusion phenomena in many areas,
e.g., turbulence [KBS87], disordered medium [BG90], intermittent chaotic
systems [ZK93], etc. However, the kinetic equations have two advantages over
a random walk approach: firstly, they allow one to explore various boundary
conditions (e.g., reflecting and/or absorbing) and, secondly, to study
diffusion and/or relaxation phenomena in external fields. Both possibilities
are difficult to realize in the framework of CTRW.

There are three types of fractional kinetic equations: the first one,
describing Markovian processes, contains equations with fractional space or
velocity derivative, the second one, describing non-Markovian processes,
contains equations with fractional time derivative, and the third class,
naturally, contains both fractional space and time derivatives, as well.
However, all three types are suitable to describe time evolution of the
probability density function (PDF) of a very narrow class of diffusion
processes, which are characterized by a unique diffusion exponent showing
time-dependence of the characteristic displacement (e.g., of the root mean
square) [MK00]. These processes are also called fractal, or self-affine
processes, and they are characterized by the exponent $H$, called the Hurst
exponent, which depends on the order of fractional derivative in the kinetic
equation. We recall that the stochastic process $x(t)$ is self-affine, or
fractal, if its stationary increments possess the following property [ST94]: 
\begin{equation}
x(t+\kappa \tau )-x(t)\mathop =\limits^d\kappa ^H(x(t+\tau )-x(t))  \tag{1}
\end{equation}
where $\kappa $ and $H$ are positive constants. The sign $\mathop  %
=\limits^d $implies, that the left and the right hand sides of Eq.(1) have
the same PDFs.

As a possible generalization of fractional kinetic equations, we propose
fractional diffusion equations in which the fractional order derivatives are
integrated with respect to the order of differentiation (\textit{distributed
order fractional diffusion equations}). They can serve as a paradigm for the
kinetic description of the random processes possessing non-unique diffusion
exponent and hence, non-unique Hurst exponent. The processes with
time-dependent Hurst exponent are believed to provide useful models for a
host of continuous and non-stationary natural signals; they are also
constructed explicitly [PV95], [AV99], [AV00]. Ordinary differential
equations with distributed order derivatives were proposed in the works by
Caputo [Cap69], [Cap95] for generalizing stress-strain relation of unelastic
media. In Refs. [BT00], [BT00a] the method of the solution was proposed
which is based on generalized Taylor series representation. A basic
framework for the numerical solution of distributed order differential
equations was introduced in [DF01]. Very recently, Caputo [Cap01] proposed
the generalization of the Fick's law using distributed order time derivative.

We write the \textit{distributed order time fractional diffusion equation}
for the PDF $f(x,t)$ as 
\begin{equation}
\int\limits_0^1{d\beta \tau ^{\beta -1}p(\beta )}\frac{{\partial ^\beta f}}{{%
\partial t^\beta }}=D\frac{{\partial ^2f}}{{\partial x^2}}%
\,\,,\,\,\,\,\,\,\,f(x,0)=\delta (x),  \tag{2}
\end{equation}
where $\tau $ and $D$ are positive constants, [$\tau $] = sec, [$D$] =
cm2/sec, $p(\beta )$ is a dimensionless non- negative function of the order
of the derivative, and the time fractional derivative of order $\beta $ is
understood in the Caputo sense [GM97]: 
\begin{equation}
\frac{{\partial ^\beta f}}{{\partial t^\beta }}=\frac 1{{\Gamma (1-\beta )}%
}\int\limits_0^t{d\tau (t-\tau )^{-\beta }\frac{{\partial f}}{{\partial t}}}
\tag{3}
\end{equation}
If we set $p(\beta )=\delta (\beta -\beta _0)$, $0<\beta _0\leq 1$, then we
arrive at time fractional diffusion equation, whose solution is the PDF of
the self-affine random process with the Hurst exponent equal to $\beta _0/2$%
. The PDF is expressed through Wright function [Mai97]. The diffusion
process is then characterized by the mean square displacement 
\begin{equation}
\left\langle {x^2}\right\rangle (t)\equiv \int\limits_{-\infty }^\infty {%
dxx^2f(x,t)=}\frac 2{{\Gamma \left( {\beta _0+1}\right) }}D\tau ^{1-\beta
_0}t^{\beta _0}.  \tag{4}
\end{equation}
This formula provides the generalization of the corresponding formula for
classical diffusion valid at $\beta _0$=1. Since $\beta _0$ can be less than
1, Eq.(4) describes the process of slow diffusion, or sub- diffusion.

Let us now prove that the solution of Eq.(2) is a PDF. The derivation here
parallels to the method used in [Sok01]. Its aim is to show that the random
process whose PDF obeys Eq.(2) is subordinated to the Wiener process.
Returning to Eq.(2) and applying the transformations of Laplace and Fourier
in succession, 
\begin{equation}
\stackrel{\stackrel{\wedge }{\backsim }}{f}(k,s)=\int\limits_{-\infty
}^{\infty }{dxe^{ikx}\int\limits_{0}^{t}{dte^{-st}}f(x,t)\,\,,\,}  \tag{5}
\end{equation}
we get from Eq.(2), 
\begin{equation}
\stackrel{\stackrel{\wedge }{\backsim }}{f}(k,s)=\frac{1}{s}\frac{{I(s\tau )}%
}{{I(s\tau )+k^{2}D\tau }}\,\,,  \tag{6}
\end{equation}
where 
\begin{equation}
I(s\tau )=\int\limits_{0}^{1}{d\beta }(s\tau )^{\beta }p(\beta )  \tag{7}
\end{equation}
We note that under the conditions described above the function $I(s\tau )$
is completely monotone on the positive real axis, i.e., it is positive and
the signs of its derivatives alternate. We rewrite Eq.(6) as follows:

\begin{equation}
\stackrel{\stackrel{\wedge }{\backsim }}{f}(k,s)=\frac{I}{s}%
\int\limits_{0}^{\infty }{due^{-u\left[ {I+k^{2}D\tau }\right]
}=\int\limits_{0}^{\infty }{due^{-uk^{2}D\tau }\tilde{G}(u,s)\,\,,}}  \tag{8}
\end{equation}
where 
\begin{equation}
\tilde{G}(u,s)=\frac{{I(s\tau )}}{s}e^{-uI(s\tau )}  \tag{9}
\end{equation}
is the Laplace transform of a function $G(u,t)$ whose properties will be
specified below. Now, $f(x,t)$ can be written as 
\begin{equation}
f(x,t)=\int\limits_{-\infty }^{\infty }{\frac{{dk}}{{2\pi }}%
e^{-ikx}\int\limits_{Br}{\frac{{ds}}{{2\pi i}}e^{st}\int\limits_{0}^{\infty }%
{due^{-uk^{2}D\tau }}}}\tilde{G}(u,s)=\int\limits_{0}^{\infty }{du\frac{{%
e^{-x^{2}/4uD\tau }}}{{\sqrt{4\pi uD\tau }}}G(u,t)\,.}  \tag{10}
\end{equation}
The function $G(u,t)$ is the PDF providing the subordination transformation,
from time scale $t$ to time scale $u$. Indeed, at first we note that $G(u,t)$
is normalized with respect to $u$ for any $t$. With the use of Eq.(9) we get 
\begin{equation}
\int\limits_{0}^{\infty }{duG(u,t)=\mathbf{L}_{s}^{-1}\int\limits_{0}^{%
\infty }{du\left[ {\frac{I}{s}e^{-uI}}\right] }}=\mathbf{L}_{s}^{-1}\left[ {%
\frac{1}{s}}\right] =1\,\,,  \tag{11}
\end{equation}
where $\mathbf{L}_{s}^{-1}$ is an inverse Laplace transformation. Now, to
prove the positivity of $G(u,t)$, it is sufficient to show that its Laplace
transform $\tilde{G}(u,s)$ is completely monotone on the positive real axis
[Fel71]. The last statement arises from the observation that $\tilde{G}(u,s)$
is a product of two completely monotone functions, $I/s$ and $exp(-uI)$. The
monotonicity of the former is obvious, whereas the monotonicity of the
latter is an elementary consequence of the Criterion 2 in [Fel71], Chapter
XIII, Section 4. Thus, we may conclude that the solution of Eq.(2) is a PDF,
and that the random process, whose PDF obeys a distributed order time
fractional diffusion equation, is subordinated to the Gaussian process using
operational time.

Since we are interested in diffusion problem, there is no need to obtain
PDF, but its second moment only. Using Eq.(6) we get 
\begin{equation}
\left\langle {x^{2}}\right\rangle (t)=\left. {\left\{ {\ -\frac{{\partial
^{2}\hat{f}(k,t)}}{{\partial k^{2}}}}\right\} }\right| _{k=0}=2D\tau \mathbf{%
L}_{s}^{-1}\left\{ {\frac{1}{{sI(s\tau )}}}\right\} .  \tag{12}
\end{equation}
Consider two fractional exponents in Eq.(2), namely, let 
\begin{equation}
p(\beta )=B_{1}\delta (\beta -\beta _{1})+B_{2}\delta (\beta -\beta
_{2})\,\,,  \tag{13}
\end{equation}
where $0<\beta _{1}<\beta _{2}\leq 1$, $B1>0$, $B2>0$. Inserting Eq.(13)
into Eq.(12) we get, denoting $b_{1}=B_{1}\tau ^{\beta _{1}},b_{2}=B_{2}\tau
^{\beta _{2}}\,:$ 
\begin{equation}
\left\langle {x^{2}}\right\rangle (t)=2D\tau \mathbf{L}_{s}^{-1}\left\{ {%
\frac{1}{{s\left( {b_{1}s^{\beta _{1}}+b_{2}s^{\beta _{2}}}\right) }}}%
\right\} =\frac{{2D\tau }}{{b_{2}}}\mathbf{L}_{s}^{-1}\left\{ {\frac{{%
s^{-\beta _{1}-1}}}{{\frac{{b_{1}}}{{b_{2}}}+s^{\beta _{2}-\beta _{1}}}}}%
\right\} .  \tag{14}
\end{equation}
Recalling the Laplace transform of the generalized Mittag-Leffler function $%
E_{\mu ,\nu }(z),\,\,\mu >0,\,\,\nu >0$, which can be conveniently written
as [GM97] 
\begin{equation}
\mathbf{L}_{t}\left\{ {t^{\nu }E_{\mu ,\nu }(-\lambda t^{\mu }}\right\} =%
\frac{{s^{\mu -\nu }}}{{s^{\mu }+\lambda }},\quad {\mathop{\rm Re}}s>\left|
\lambda \right| ^{1/\mu },  \tag{15}
\end{equation}
we get from Eq.(14), 
\begin{equation}
\left\langle {x^{2}}\right\rangle =\frac{{2D\tau }}{{b_{2}}}t^{\beta
_{2}}E_{\beta _{2}-\beta _{1},\beta _{2}+1}\left( {\ -\frac{{b_{1}}}{{b_{2}}}%
t^{\beta _{2}-\beta _{1}}}\right) .  \tag{16}
\end{equation}
To get asymptotics at small $t$, we use an expansion, which is, in fact the
definition of $E_{\mu ,\nu }(z)$, see [Erd55], Ch.XVIII, Eq.(19): 
\begin{equation}
E_{\mu ,\nu }(z)=\sum\limits_{n=0}^{\infty }{\frac{{z^{n}}}{{\Gamma (\mu
n+\nu )}},}  \tag{17}
\end{equation}
which yields in the main order for the square displacement , 
\begin{equation}
\left\langle {x^{2}}\right\rangle \approx \frac{{2D\tau }}{{B_{2}\Gamma
(\beta _{2}+1)}}\left( {\frac{t}{\tau }}\right) ^{\beta _{2}}\propto
t^{\beta _{2}}.  \tag{18}
\end{equation}
For large $t$ we use the following expansion valid on the real negative
axis, see [Erd55], Ch.XVIII, Eq.(21), 
\begin{equation}
E_{\mu ,\nu }(z)=-\sum\limits_{n=1}^{N}{\frac{{z^{-n}}}{{\Gamma (-\mu n+\nu )%
}}+O\left( {\left| z\right| ^{-1-N}}\right) ,\left| z\right| \to \infty ,} 
\tag{19}
\end{equation}
which yields 
\begin{equation}
\left\langle {x^{2}}\right\rangle \approx \frac{{2D\tau }}{{B_{1}\Gamma
(\beta _{1}+1)}}\left( {\frac{t}{\tau }}\right) ^{\beta _{1}}\propto
t^{\beta _{1}}.  \tag{20}
\end{equation}
Since $\beta _{1}<\beta _{2}$, we have the effect of \textit{diffusion with
retardation}. We also note that the kinetic equation with two fractional
derivatives of different orders appears quite naturally when describing
subdiffusive motion in velocity fields [MKS98]. In this case the order of
derivatives are $\beta $ and $\beta -1$, so that the situation differs from
one discussed above.

Now we consider a simple particular case which, in some sense, is opposite
to the cases considered above. Namely, we put 
\begin{equation}
p(\beta )=1,0\le \beta \le 1.  \tag{21}
\end{equation}
Inserting Eq.(21) into Eq.(7) we get 
\begin{equation}
I(s\tau )=\frac{{s\tau -1}}{{\log (s\tau )}},  \tag{22}
\end{equation}
and, using then Eq.(12), 
\begin{equation}
\left\langle {x^2}\right\rangle =2D\tau \left\{ {\log \frac t\tau +\gamma
+e^{t/\tau }E_1\left( {\frac t\tau }\right) }\right\} ,  \tag{23}
\end{equation}
where $\gamma =0.5772$: is the Eiler constant, and 
\begin{equation}
E_1(z)=\int\limits_z^\infty {dy\frac{{e^{-y}}}y}  \tag{24}
\end{equation}
is the exponential integral. Using now the expansions valid on the positive
real axis, see [AS65], Eqs.(5.1.11) and (5.1.51), respectively, 
\begin{equation}
E_1(z)=-\gamma -\log z-\sum\limits_{n=1}^\infty {\frac{{(-z)^n}}{{n\cdot n!}}%
,\,\,\,z\rightarrow 0,}  \tag{25}
\end{equation}
and 
\begin{equation}
E_1(z)\approx \frac{{e^{-z}}}z\sum\limits_{n=0}^N{(-1)^n\frac{{n!}}{{z^n}}%
,\,\,z\rightarrow \infty ,}  \tag{26}
\end{equation}
we get, retaining the main terms of the asymptotics at small and large
times, respectively: 
\begin{equation}
\left\langle {x^2}\right\rangle \approx \left\{ {\begin{array}{*{20}c}
{2D\tau \frac{t}{\tau }\log \frac{\tau }{t},t \to 0} \\ {2D\tau \log \left(
{\frac{t}{\tau }} \right),t \to \infty } \\ \end{array}}\right.  \tag{27}
\end{equation}
Thus, at small times we have slightly anomalous super-diffusion, whereas at
large times we have \textit{superslow diffusion}.

The formula (27) can be generalized to the case 
\begin{equation}
p(\beta )=\left\{ {\begin{array}{*{20}c} {\frac{1}{{\beta _2 - \beta _1 }},0
\le \beta _1 \le \beta \le \beta _2 \le 1} \\ {0,otherwise} \\ \end{array}}%
\right. ,  \tag{28}
\end{equation}
and $\int\limits_0^\infty {d\beta p(\beta )=1}$. Inserting Eq.(28) into
Eq.(7) and, then into Eq.(12), we get 
\begin{eqnarray}
\left\langle {x^2}\right\rangle (t) &=&\frac{{2D\tau }}{{\beta _2-\beta _1}}%
\mathbf{L}_s^{-1}\left\{ {\frac{{\log (s\tau )}}{{s\left[ {(s\tau )^{\beta
_2}-(s\tau )^{\beta _1}}\right] }}}\right\} =  \tag{29} \\
&=&-\frac{{2D\tau }}{{\beta _2-\beta _1}}\left. {\left\{ {\frac d{{d\delta }}%
\mathbf{L}_s^{-1}\left[ {\frac{{s^{-\delta -\beta _1}}}{{s^{\beta _2-\beta
_1}-1}}}\right] }\right\} \left( {\frac t\tau }\right) }\right| _{\delta =1}
\nonumber
\end{eqnarray}
Recalling Eq.(15) mean square displacement can be written as 
\[
\left\langle {x^2}\right\rangle =\frac{{2D\tau }}{{\beta _2-\beta _1}}\left( 
{\frac t\tau }\right) ^{\beta _2}\left\{ {\log \left( {\frac \tau t}\right)
E_{\beta _2-\beta _1,\beta _2+1}\left( {\left( {\frac t\tau }\right) ^{\beta
_2-\beta _1}}\right) -}\right. 
\]

\begin{equation}
-\left. {\left. {\left( {\frac d{{d\delta }}E_{\beta _2-\beta _1,\beta
_2+\delta }\left( {\left( {\frac t\tau }\right) ^{\beta _2-\beta _1}}\right) 
}\right) }\right| _{\delta =1}}\right\} .  \tag{30}
\end{equation}
Using Eq.(17), we get an expansion for $\left\langle {x^2}\right\rangle $ at
small $t$: 
\[
\left\langle {x^2}\right\rangle =\frac{{2D\tau }}{{\beta _2-\beta _1}}\left( 
{\frac t\tau }\right) ^{\beta _2}\sum\limits_{n=0}^\infty {\left( {\frac
t\tau }\right) }^{n\left( {\beta _2-\beta _1}\right) }\left\{ {\log \left( {%
\frac \tau t}\right) +}\right. 
\]

\begin{equation}
\left. {\ +\psi \left( {1+\beta _2+n\left( {\beta _2-\beta _1}\right) }%
\right) }\right\} \Gamma ^{-1}\left( {1+\beta _2+n\left( {\beta _2-\beta _1}%
\right) }\right)  \tag{31}
\end{equation}
where 
\[
\psi (\nu )=d\left( {\log \Gamma (\nu )}\right) /d\nu 
\]
is the $\psi $ - function. At large $t$ we explore the asymptotics valid on
the real positive axis see [Erd55], Ch.XVIII, Eq.(22), 
\begin{equation}
E_{\mu ,\nu }(z)=\frac 1\mu z^{\left( {1-\nu }\right) /\mu }\exp \left( {%
z^{1/\mu }}\right) -\sum\limits_{n=1}^N{\frac{{z^{-n}}}{{\Gamma \left( {\nu
-\mu n}\right) }}+O\left( {\left| z\right| ^{-1-N}}\right) .}  \tag{32}
\end{equation}
With using Eq.(32), Eq.(30) takes on the form 
\[
\left\langle {x^2}\right\rangle \approx \frac{{2D\tau }}{{\beta _2-\beta _1}}%
\left( {\frac t\tau }\right) ^{\beta _1}\sum\limits_{n=0}^N{\left( {\frac
t\tau }\right) }^{-n\left( {\beta _2-\beta _1}\right) }\{\log \left( {\frac
t\tau }\right) - 
\]

\begin{equation}
-\psi \left( {1+\beta _2-n\left( {\beta _2-\beta _1}\right) }\right)
\}\Gamma ^{-1}\left( {1+\beta _2-n\left( {\beta _2-\beta _1}\right) }\right)
.  \tag{33}
\end{equation}
If we set $\beta _1=0,\,\,\beta _2=1$ in Eqs.(31) and (33), then we arrive
at the same expansions which are obtained by inserting Eqs.(25) and (26)
into Eq.(23), respectively. In particular, the main term of the series (31)
and (33) at $\beta _1=0,\,\,\beta _2=1$ coincide with Eq.(27).

The fractional diffusion equations with a given order of fractional time
derivative are closely connected to the continuous-time random walk
processes (CTRW) with the power-law distribution of waiting times between
the subsequent steps [MK00], [BMK00]. Now we establish the connection
between the distributed order time fractional diffusion equations and more
general CTRW situations. Recall the basic formula of the CTRW in the
Fourier-Laplace space [KBS87]:

\begin{equation}
\stackrel{\stackrel{\wedge }{\backsim }}{f}(k,s)=\frac{{1-\tilde{w}(s)}}{s}%
\frac{1}{{1-\stackrel{\stackrel{\wedge }{\backsim }}{\psi }(k,s)}},  \tag{34}
\end{equation}
where $\tilde{w}(s)$ is the Laplace transform of the waiting-time PDF $w(t)$%
, and $\stackrel{\stackrel{\wedge }{\backsim }}{\psi }(k,s)$ is the
Fourier-Laplace transform of the joint PDF of jumps and waiting times $\psi
(\xi ,t)$. Assume the decoupled joint PDF, $\psi (\xi ,t)=\lambda \left( \xi
\right) w(t)$, and that the jump length variance is finite, that is, the
Fourier transform of $\lambda \left( \xi \right) $ is 
\begin{equation}
\hat{\lambda}(k)\approx 1-D\tau k^{2}  \tag{35}
\end{equation}
to the lowest orders in $k$. Then, we consider the situations, in which mean
waiting time diverges, that is, at large $t$ the waiting time PDF behaves as 
\begin{equation}
w(t)\approx \tau ^{\beta }/t^{1+\beta },\,\,\,0<\beta <1,  \tag{36}
\end{equation}
and, consequently, 
\begin{equation}
\tilde{w}(s)\approx 1-(s\tau )^{\beta }  \tag{37}
\end{equation}
at small $s$. If $\beta $ is constant, then inserting Eqs.(37) and (35) into
Eq.(34) and making an inverse Fourier-Laplace transform, we arrive at time
fractional diffusion equation. Now let us consider the case when $\beta $ 
\textit{fluctuates}. Indeed, for example, in the model called the Arrhenius
cascade, which is inspired from studies of disordered systems, the unique $%
\beta $ appears only under the assumption that the random trapping time is
related to the random height of the well by the Arrhenius law [Bar99]. In a
more realistic model, this law gives only the average value of the trapping
time. Thus, we may speculate that in order to take into account the
fluctuations of the trapping time, we can introduce the conditional PDF 
\begin{equation}
w(t|\beta )\approx \tau ^{\beta }/t^{1+\beta },  \tag{38}
\end{equation}
and the PDF $p(\beta )$, as well. Now, we have the relation 
\begin{equation}
w(t)=\int\limits_{0}^{1}{d\beta p(\beta )w(t;\beta ),}  \tag{39}
\end{equation}
where [0;1] is the whole interval for variations of $\beta $. We note that
all waiting-time distributions with $\beta \geq 1$ correspond to similar
behavior described by the first order derivative. Then, for the $\tilde{w}(s)
$ we have, instead of Eq.(37), 
\begin{equation}
\tilde{w}(s)\approx 1-\int\limits_{0}^{1}{d\beta }(s\tau )^{\beta }p(\beta
),\,\,\,\,\,\,\,\,\,\,\,p(\beta )\ge 0,\,\,\,\int\limits_{0}^{1}{d\beta
p(\beta )=1.}\,  \tag{40}
\end{equation}
Inserting Eqs.(40) and (35) into Eq.(34) we arrive at Eqs.(6) and (7). Thus,
we see that the weight function $p(\beta )$ has the meaning of the PDF.

The model with fluctuating $\beta $ is, of course, only one of the possible
interpretations of Eq.(39): the non-exact power-law behavior of the
waiting-time PDF can physically have very different reasons. In particular,
the representation (39) allows us to consider \textit{regularly varying}
waiting- time PDFs, i.e., those which behave at $t\rightarrow \infty $ as $%
w(t)\propto t^{-1-\beta }g(t)$ , where $g(t)$ is a slowly varying function,
e.g., any power of $log\,\,t$ [Fel71]. We are also able to consider a
waiting-time PDFs $w(t)$ which show an approximately scaling behavior with
the exponents \textit{changing with time}. For such distributions the
effective PDFs $p(\beta )$ can be determined, and thus such non-perfectly
scaling CTRWs can be described through distributed-order diffusion
equations. The formal inversion of Eq.(39) can follow by noting that $tw(t)$
taken as a function of $log\,\,t$ is the Laplace-transform of the function

\[
\phi (\beta )=\left\{ {%
\begin{array}{*{20}c}
   {\tau ^\beta  p(\beta ),0 \le \beta  \le 1}  \\
   {0,1 < \beta  < \infty }  \\
\end{array}}\right. . 
\]
Indeed, 
\[
tw(t)=\int\limits_0^\infty {d\beta \phi (\beta )t^{-\beta
}=\int\limits_0^\infty {d\beta \phi (\beta )\exp \left( {\ -\beta \log t}%
\right) =\mathbf{L}_\beta \left\{ {\phi (\beta )}\right\} (\log t).}} 
\]
The function $\phi (\beta )$ is thus given by 
\[
\phi (\beta )=\mathbf{L}_u^{-1}\left\{ {e^uw(e^u)}\right\} . 
\]
The value of $\tau $ can then be found through the normalization condition 
\[
\int\limits_0^\infty {\phi (\beta )}\tau ^{-\beta }=1, 
\]
which defines then the function $p(\beta )$. The description of the process
through distributed-order diffusion equation is possible whenever this
function is non-negative and concentrated on $0\leq \beta \leq 1$.

Now we turn to another type of fractional equation, namely, \textit{%
distributed order space fractional diffusion equation} which, in dimensional
variables, takes on the form 
\begin{equation}
\frac{{\partial f}}{{\partial t}}=\int\limits_{0^{+}}^2{d\alpha D(\alpha )}%
\frac{{\partial ^\alpha f}}{{\partial \left| x\right| ^{^\alpha }}}%
,f(x,0)=\delta (x),  \tag{41}
\end{equation}
where $D$ is a (dimensional) function of the order of the derivative $\alpha 
$, and the Riesz space fractional derivative $\partial ^\alpha /\partial
\left| x\right| ^\alpha $ is understood through its Fourier transform $\hat{%
\Phi}$ as 
\begin{equation}
\hat{\Phi}\left( {\frac{{\partial ^\alpha f}}{{\partial \left| x\right|
^\alpha }}}\right) \div -\left| k\right| ^\alpha \hat{f}.  \tag{42}
\end{equation}
If we set $D(\alpha )=K_{\alpha _0}\delta (\alpha -\alpha _0)$, then we
arrive at the space fractional diffusion equation, whose solution is a Levy
stable PDF of the self-affing stable process whose Hurst exponent is equal $%
1/\alpha _0$. The PDF is expressed in terms of the Fox$^{\prime }$s
H-function [Fox61], [Sch86].\thinspace In the general case $D\left( \alpha
\right) $ can be represented as 
\begin{equation}
D(\alpha )=l^{\alpha -2}DA(\alpha ),  \tag{43}
\end{equation}
where $l$ and $D$ are dimensional positive constants, [$l$] = cm, [$D$] = cm$%
^2$/sec, $A$ is a dimensionless non-negative function of $\alpha $. The
equation which follows for the characteristic function from Eq.(41) has the
solution 
\begin{equation}
\hat{f}(k,t)=\exp \left\{ {\ -\frac{{Dt}}{{l^2}}\int\limits_0^2{d\alpha A}%
(\alpha )(\left| k\right| l)^\alpha }\right\} .  \tag{44}
\end{equation}
Note that the normalization condition, 
\begin{equation}
\int\limits_{-\infty }^\infty {dxf(x,t)=\hat{f}(k=0,t)=1,}  \tag{45}
\end{equation}
is fulfilled.

Consider the simple particular case, 
\begin{equation}
A(\alpha )=A_1\delta (\alpha -\alpha _1)+A_2\delta (\alpha -\alpha _2), 
\tag{46}
\end{equation}
where $0<\alpha _1<\alpha _2,\,\,A1>0,\,\,A2>0$. Inserting Eq.(46) into
Eq.(44) we have 
\begin{equation}
\hat{f}(k,t)=\exp \left\{ {\ -a_1\left| k\right| ^{\alpha _1}t-a_2\left|
k\right| ^{\alpha _2}t}\right\} ,  \tag{47}
\end{equation}
where $a_1=A_1D/l^{2-\alpha _1},a_2=A_2D/l^{2-\alpha _2}.$ The
characteristic function (47) is the product of two characteristic functions
of the Levy stable PDFs with the Levy indexes $\alpha _1,\,\alpha _2$, and
the scale parameters $a_1^{1/\alpha _1}$ and $a_2^{1/\alpha _2}$,
respectively. Therefore, the inverse Fourier transformation of Eq.(58) gives
the PDF which is the convolution of the two stable PDFs, 
\begin{equation}
f(x,t)=a_1^{-1/\alpha _1}a_2^{-1/\alpha _2}t^{-1/\alpha _1-1/\alpha
_2}\int\limits_{-\infty }^\infty {dx^{\prime }\mathit{L}_{\alpha _1,0}\left( 
{\frac{{x-x^{\prime }}}{{(a_1t)^{1/\alpha _1}}}}\right) \mathit{L}_{\alpha
_2,0}\left( {\frac{{x^{\prime }}}{{(a_2t)^{1/\alpha _2}}}}\right) ,} 
\tag{48}
\end{equation}
where $\mathit{L}_\alpha \left( x\right) $ is the PDF of the symmetric Levy
stable law possessing the characteristic function 
\begin{equation}
\mathit{\hat{L}}_{\alpha ,0}(k)=\exp \left( {\ -|k|^\alpha }\right) . 
\tag{49}
\end{equation}
The PDF given by Eq.(48) is, obviously, positive, as the convolution of two
positive PDFs. The PDF will be also positive, if the function $A(\alpha )$
is represented as a sum of $N$ delta-functions multipolied by positive
constants, $N$ is a positive integer. Moreover, if $A(\alpha )$ is a
continuos positive function, then discretizing the integral in Eq.(41) by a
Riemann sum and passing to the limit we can also conclude on the positivity
of the PDF.

Since the mean square displacement diverges for the Levy stable process, the
anomalous superdiffusion can be characterized by the typical displacement $%
\delta x$ of the diffusing particle [WS82], 
\begin{equation}
\delta x\propto \left\langle {\left| x\right| ^q}\right\rangle ^{1/q}, 
\tag{50}
\end{equation}
where $\left\langle {\left| x\right| ^q}\right\rangle $ is the $q$-th
absolute moment of the PDF obeying Eq.(41). For the stable process with the
Levy index $\alpha $ 
\begin{equation}
\left\langle {\left| x\right| ^q}\right\rangle =\left\{ {%
\begin{array}{*{20}c} {C(q;\alpha )t^{q/\alpha } ,0 < q < \alpha < 2} \\
{\infty ,q \ge \alpha } \\ \end{array}}\right. ,  \tag{51}
\end{equation}
where the coefficient

\begin{equation}
C(q;\alpha )=\frac 2{{\pi q}}\left( {K_\alpha t}\right) ^{q/\alpha }\sin
\left( {\frac{{\pi q}}2}\right) \Gamma (1+q)\Gamma \left( {1-\frac q\alpha }%
\right)  \tag{52}
\end{equation}
was obtained in [WS82]. To evaluate the $q$-th moment for the case given by
Eq.(46), $q<\alpha _1$, we use the following expression , see, e.g., [Zol86] 
\begin{equation}
\left\langle {\left| x\right| ^q}\right\rangle =\frac 2\pi \Gamma (1+q)\sin
\left( {\frac{{\pi q}}2}\right) \int\limits_0^\infty {dk\left( {1-{%
\mathop{\rm Re}\nolimits}\hat{f}(k,t)}\right) }k^{-q-1}.  \tag{53}
\end{equation}
We insert Eq.(47) into Eq.(53) and expand in series either $\exp (-a_1\left|
k\right| ^{\alpha _1}t)$, or $\exp (-a_2\left| k\right| ^{\alpha _2}t)$,
with subsequent integration over $k$. As the result, for the $q$-th moment
we have expansions valid at $q<\alpha _1$ and for small and large times,
respectively,

\[
\left\langle {\left| x \right|^q } \right\rangle = \frac{2}{{\pi q}}\left( {%
a_2 t} \right)^{q/\alpha _2 } \sin \left( {\frac{{\pi q}}{2}} \right)\Gamma
(1 + q)\Gamma \left( {1 - \frac{q}{{\alpha _2 }}} \right) \times 
\]

\begin{equation}
\times \left\{ {1+\frac q{{\Gamma \left( {1-\frac q{{\alpha _2}}}\right) }%
}\sum\limits_{n=1}^\infty {\frac{{(-1)^{n+1}}}{{\alpha _2n!}}%
a_1^na_2^{-n\alpha _1/\alpha _2}\Gamma \left( {\frac{{n\alpha _1-q}}{{\alpha
_2}}}\right) t^{n\left( {1-\alpha _1/\alpha _2}\right) }}}\right\}
,\,\,t\rightarrow 0,  \tag{54}
\end{equation}

\[
\left\langle {\left| x \right|^q } \right\rangle \approx \frac{2}{{\pi q}}%
\left( {a_1 t} \right)^{q/\alpha _1 } \sin \left( {\frac{{\pi q}}{2}}
\right)\Gamma (1 + q)\Gamma \left( {1 - \frac{q}{{\alpha _1 }}} \right)
\times 
\]

\begin{equation}
\left\{ {1+\frac q{{\Gamma \left( {1-\frac q{{\alpha _1}}}\right) }%
}\sum\limits_{n=1}^N{\frac{{(-1)^{n+1}}}{{\alpha _1n!}}a_2^na_1^{-n\alpha
_2/\alpha _1}\Gamma \left( {\frac{{n\alpha _2-q}}{{\alpha _1}}}\right)
t^{-n\left( {\alpha _2/\alpha _1-1}\right) }}}\right\} ,\,\,t\rightarrow
\infty ,  \tag{55}
\end{equation}
One can see, that at small times the characteristic displacement grows as $%
t^{1/\alpha _2}$, whereas at large times it grows as $t^{1/\alpha _1}$.
Thus, we have \textit{superdiffusion with acceleration}.

In summary, we believe that the distributed order fractional diffusion
equations can serve as a useful tool for the description of complicated
diffusion processes, for which the diffusion exponent can change in the
course of time. Further investigations are needed to establish the
connection between proposed kinetics and multifractality. On the other hand,
the development of numerical schemes for solving distributed order kinetic
equations and for modelling sample paths of the random processes governed by
these equations is also of importance

.

{\large Acknowledgements}

This work is supported by the INTAS Project 00 - 0847. AC also acknowledges
the hospitality of the Department of Mathematics and Computer Science of
Free University of Berlin and support from the DAAD.

\medskip

{\large References}

\medskip

[AS65] M. Abramowitz, I.A. Stegun. \textit{Handbook of Mathematical Functions%
}, Dover, New York, 1965.

[AV99] A. Ayache, J. Levy Vehel, Generalized Multifractional Brownian Motion
: Definition and Preliminary Results. In : M. Dekking, J. Levy Vehel, E.
Lutton and C. Triccot (eds) \textit{Fractals : Theory and Applications in
Engineering}. Springer-Verlag 1999, p.17-32.

[AV00] A. Ayache, J. Levy Vehel, The Generalized Multifractional Brownian
Motion. \textit{Statistical Inference for Stochastic Processes}, vol.1-2,
7-18 (2000).

[Bar99] F. Bardou, Cooling gases with Levy flights: using the generalized
central limit theorem in physics.In: O. E. Barndorff-Nielsen, S. E.
Graversen and T. Mikosch (eds.), Proceedings of the Conference "Levy
processes. Theory and applications." University of Aarhus, 18-22 January
1999, pp.22-38 (downloadable from
http://www.maphysto.dk/publications/MPS-misc/1999/11.pdf)

[BG90] J.-P. Bouchaud, A. Georges, Anomalous Diffusion in Disordered Media:
Statistical Mechanisms, Models and Physical Applications. \textit{Phys.
Reports} \textbf{195}, Nos.4-5, 127-293 (1990).

[BMK00] E. Barkai, R. Metzler, J. Klafter, From continuous time random walks
to the fractional Fokker-Planck equation. \textit{Physical Review} \textbf{%
E61}, No.1, 132-138 (2000).

[BT00] R. L. Bagley, P.J. Torvik, On the Existence of the Order Domain and
the Solution of Distributed Order Equations. - Part I. \textit{International
Journal of Applied Mathematics}, vol.2, No.7, 865-882 (2000).

[BT00a] R. L. Bagley, P.J. Torvik, On the Existence of the Order Domain and
the Solution of Distributed Order Equations. - Part II. \textit{%
International Journal of Applied Mathematics}, vol.2, No.8, 965-987 (2000).

[Cap69] M. Caputo, \textit{Elasticita e dissipazione}, Zanichelli Printer,
Bologna (1969).

[Cap95] M. Caputo, Mean Fractional Order Derivatives. Differential Equations
and Filters. \textit{Annalls Univ. Ferrara} - Sez. VII - Sc. Mat., Vol. XLI,
73-84 (1995).

[Cap01] M. Caputo, Distributed Order Differential Equations Modelling
Dielectric Induction and Diffusion. \textit{Fractional Calculus and Applied
Analysis}, Vol.4, No.4, 421-442 (2001).

[Com96] A. Compte, Stochastic foundations of fractional dynamics. \textit{%
Physical Review} \textbf{E53}, No.4, 4191 - 4193 (1996).

[DF01] K. Diethelm, N. J. Ford, Numerical Solution Methods for Distributed
Order Differential Equations. \textit{Fractional Calculus and Applied
Analysis}, vol.4, No.4, 531-542 (2001).

[Erd55] A. Erdelyi, ed. \textit{Bateman Manuscript Project. Higher
Transcendental Functions}, vol.III. McGraw-Hill, Inc., New York, 1955.

[Fel71] W. Feller, \textit{An Introduction to Probability Theory and Its
Applications}. John Wiley and Sons, Inc., New York, 2-nd ed., 1971, vol.II.

[Fox61] C. Fox, The G and H functions as symmetrical Fourier kernels. 
\textit{Transactions of the American Mathematical Society} \textbf{98},
No.3, 395-429 (1961).

[GM97] R. Gorenflo, F. Mainardi, in A. Carpinteri, F. Mainardi (eds.) 
\textit{Fractals and Fractional Calculus in Continuum Mechanics}, Springer
Verlag, Wien and New York 1997, p.223 - 276.

[HA95] R. Hilfer, L. Anton, Fractional master equations and fractal time
random walks. \textit{Physical Review} \textbf{E51}, No.2, R848-R851 (1995).

[KBS87] J. Klafter, A. Blumen, M. F. Schlesinger, Stochastic pathway to
anomalous diffusion. \textit{Physical Review} \textbf{A35}, No.7, 3081-3085
(1987).

[Mai96] F. Mainardi, Fractional relaxation-oscillation and fractional
diffusion-wave phenomena. \textit{Chaos, Solitons and Fractals} \textbf{7},
1461-1477 (1996).

[Mai97] F. Mainardi, in A. Carpinteri, F. Mainardi (eds.) \textit{Fractals
and Fractional Calculus in Continuum Mechanics}, Springer Verlag, Wien and
New York 1997, p.291 - 348.

[MK00] R. Metzler and J. Klafter, The Random Walk's Guide to Anomalous
Diffusion: A Fractional Dynamics Approach. \textit{Physics Reports} \textbf{%
339}, No.1, 1-77 (2000).

[MKS98] R. Metzler, J. Klafter, I.M. Sokolov, Anomalous transport in
external fields: Continuous time random walks and fractional diffusion
equations extended. \textit{Physical Review} \textbf{E58}, No.2, 1621-1633
(1998).

[MLP01] F. Mainardi, Yu. Luchko, G. Pagnini, The Fundamental Solution of the
Space-Time Fractional Diffusion Equation. \textit{Fractional Calculus and
Applied Analysis}, vol.4, No.2, 153-192 (2001).

[MRGS00] F. Mainardi, M. Raberto, R. Gorenflo, E. Scalas, Fractional
calculus and continuous-time finance II: the waiting-time distribution. 
\textit{Physica} \textbf{A287}, No. 3-4, 468-481 (2000).

[MW65] E. W. Montroll, G. H. Weiss, Random Walk on Lattices. II. \textit{%
Journal Mathematical Physics} \textbf{6}, 167-181 (1965).

[PV95] R.-F. Peltier, J. Levy Vehel, Multifractional Brownian Motion :
Definition and Preliminary Results. INRIA Research Report No.2645, 1995. -
39 p.

[Sch86] W.R. Schneider, Stable distributions: Fox function representation
and generalization, in S. Albeverio, G. Casati, D. Merlini (eds.), \textit{%
Stochastic Processes in Classical and Quantum Systems}. Lecture Notes in
Physics, Springer Verlag, Berlin, 1986, p.497-511.

[Sok01] I.M. Sokolov, Thermodynamics and fractional Fokker-Planck equations. 
\textit{Physical Review} \textbf{E63}, 056111, 1-8.

[ST94] G. Samorodnitsky, M.S. Taqqu, \textit{Stable Non-Gaussian Processes},
Chapman and Hall, New York, 1994.

[SZ97] A. I. Saichev, G. M. Zaslavsky, Fractional kinetic equations:
solutions and applications. \textit{Chaos}, vol.7, No.4, 753-764 (1997).

[WS82] B.J. West, V. Seshadri, Linear Systems with Levy fluctuations. 
\textit{Physica} \textbf{A113}, 203 - 216 (1982).

[Zol86] V.M. Zolotarev, \textit{One-dimensional Stable Distributions},
American Math. Soc., Vol.65, (1986) [Translated from the Russian edition:
Nauka, Moscow, 1983.]

[ZK93] G. Zumofen, J. Klafter, Scale-invariant motion in intermittent
chaotic system. \textit{Physical Review} \textbf{E47}, No.2, 851-863 (1993).

\end{document}